\documentclass[pre,twocolumn]{revtex4}
\usepackage{epsfig}
\usepackage{bm}
\usepackage{amssymb}
\usepackage{amsmath}

\begin{document}

\title{Multiplicity of routes from deterministic chaos to turbulence in the flows induced by Rayleigh-Taylor instability}

\author{A. Bershadskii}

\affiliation{
ICAR, P.O. Box 31155, Jerusalem 91000, Israel
}

\begin{abstract}

The multiplicity of routes from deterministic chaos to turbulence caused by the spontaneous breaking of the local reflectional symmetry in the flows induced by Rayleigh-Taylor instability has been studied using the notion of distributed chaos. Results of numerical simulations, and laboratory and oceanic measurements have been used for this purpose. Small-scale chaotic MHD dynamo and chaotic Richtmyer-Meshkov mixing layer have been briefly discussed in this context.

\end{abstract}

\maketitle

\section{Introduction}

 The instability of an interface between fluids with different densities subject to buoyancy forces is usually referred to as Rayleigh-Taylor instability. At the nonlinear regime of this instability, bubbles/strikes of light/heavy fluid penetrating into the fluid with different density usually leave behind them regions of chaotic (turbulent) mixed fluid. The wakes of these bubbles/strikes form a broadening in time mixing zone (see Fig. 1). As with any instability this one is triggered by the initial perturbations. In this case, they are the spatial perturbations of the unstably stratified layer around the horizontal interface separating the fluids (see for a comprehensive introduction to the subject  Refs. \cite{abar}-\cite{zou2}).\\
 
 In the seminal paper Ref. \cite{hcl} three regimes of buoyancy-driven convection were introduced according to the experimental observations - chaotic (deterministic), `soft' turbulence, and `hard' turbulence. The deterministic chaos was observed for the first time in fluid dynamics just for a buoyancy-driven system \cite{lorenz}. Whereas `hard' turbulence could be related to the existence of an inertial range of scales and characterized by scaling (power) spectral law, `soft' turbulence was rather vaguely defined.  \\

  Non-smooth dynamics is usually associated with the scaling (power-law) spectra, whereas smooth dynamics can be characterized by the {\it stretched} exponential spectra. The deterministic chaos, for instance, is typically associated with smooth trajectories (sensitively dependent on the initial conditions) and characterized by {\it exponential} power spectra \cite{fm}-\cite{kds}. It is clear, however, that there can be chaotic-like dynamics different from deterministic chaos but still having {\it stretched} exponential spectrum. This type of chaotic-like dynamics can be considered as the `soft' turbulence separating the `hard' turbulence and deterministic chaos. This type of dynamics (based on the notion of distributed chaos) will be considered in the present paper. Distributed chaos is more reach than pure deterministic chaos and different intermediate regimes, with different (physically determined) stretched exponential spectra, can be observed between deterministic chaos and turbulence. \\ 
  
  We will consider different types of intermediate (distributed chaos) regimes resulting from the spontaneous breaking of local reflectional (mirror) symmetry in the flows induced by Rayleigh-Taylor instability. Results of numerical simulations, and laboratory and oceanic measurements will be used for this purpose.

\begin{figure} \vspace{-0cm}\centering \hspace{-2.1cm}
\epsfig{width=.62\textwidth,file=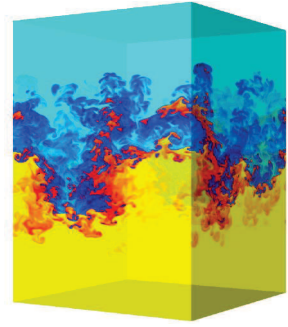} \vspace{-1.2cm}
\caption{Temperature field from a DNS \cite{bm} (a late stage of Rayleigh-Taylor mixing). 
 Blue and yellow colors correspond to heavy (cold) and light (hot) fluids, respectively. The midplane is the initial interface between the fluids. Reproduced with permission from the Annual Review of Fluid Mechanics.} 
\end{figure}

 \section{Deterministic chaos and beyond}  

   Some important properties of the Rayleigh-Taylor-induced flows can be studied in the Oberbeck-Boussinesq approximation using incompressibility and small variations of the density, due to a simple (linear) dependence of the mixture density $\rho$ on temperature $T$ (see for a recent review Ref. \cite{bm}). The system of the dynamical equations in this approximation is 
 $$
\frac{\partial {\bf u}}{\partial t} + ({\bf u} \cdot \nabla) {\bf u}  =  -\nabla p + \alpha g T {\bf e}_z + \nu \nabla^2 {\bf u} ,  \eqno{(1)}
$$
$$
\frac{\partial T}{\partial t} + ({\bf u} \cdot \nabla) T  =   \kappa \nabla^2 T, \eqno{(2)}
$$
$$
\nabla \cdot \bf u =  0, \eqno{(3)}
$$   
 ${\bf u}$ and $p$ are the velocity and pressure fields, ${\bf e}_z$ is the vertical unit vector along the gravity acceleration ${\bf g}$ , $\alpha$ is the thermal expansion coefficient, $\kappa$ and $\nu$ are the thermal and viscosity diffusivity. \\ 
 
  The initial condition, which corresponds to the Rayleigh-Taylor phenomenon, is defined by an initial perturbation on the horizontal interface separating cold (heavier) and hot (lighter) resting fluids.  The temperature jump $\theta_0$ defines the Atwood number $A = \alpha \theta_0/2 = (\rho_2-\rho_1)/( \rho_2+\rho_1)$, where  $\rho_2$ and $\rho_1$ are densities of the cold (heavy) and hot (light) fluids in their unmixed state, respectively. \\
  
  In paper Ref. \cite{ytdr} the miscible Rayleigh-Taylor instability-induced mixing was numerically studied (DNS) in a rectangular cuboid using the above-mentioned approach. For the velocity field the periodic boundary conditions were taken for the horizontal directions, whereas the bottom and top boundaries were taken as no-slip and no-flux. The no-flux boundary conditions were also taken for the temperature field. The considered dynamical time scales in this DNS were much shorter than the thermal diffusion time, hence the dynamics was dominated by the Rayleigh-Taylor instability. The separating plane interface was placed at $z = 0.06$ (in the terms of the DNS), with the spatial box restricted by $-1 \leq z \leq 1$. The parameters $\nu = \kappa = 10^{-3}$ in the terms of the DNS. Initial random temperature perturbations at the separating interface with power spectrum concentrated around wavenumber $k = 8$ were taken to trigger the instability. 
 
   The effect of the bottom and top boundaries was not significant until the mixing zone width was about $90\%$ of the vertical box size.

   Figure 2 shows the kinetic energy spectrum at the end of the DNS (the spectral data were taken from Fig. 15 of the Ref. \cite{ytdr}). The dashed curve indicates the exponential spectrum 
$$ 
E(k) \propto \exp(-k/k_c)  \eqno{(4)}
$$
  where $k_c$ (indicated by the dotted arrow in Fig. 2) is a characteristic value of the wavenumber. \\

\begin{figure} \vspace{-0.05cm}\centering
\epsfig{width=.46\textwidth,file=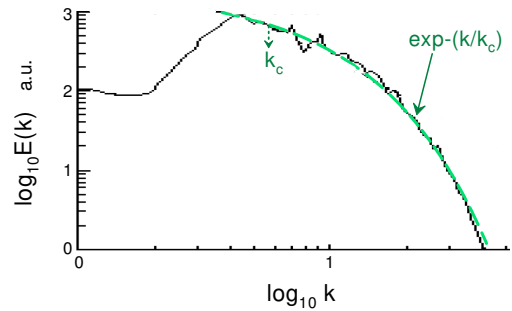} \vspace{-5.7cm}
\caption{Kinetic energy spectrum of the Rayleigh-Taylor instability-induced mixing at a chaotic stage.} 
\end{figure}
\begin{figure} \vspace{-0.5cm}\centering
\epsfig{width=.45\textwidth,file=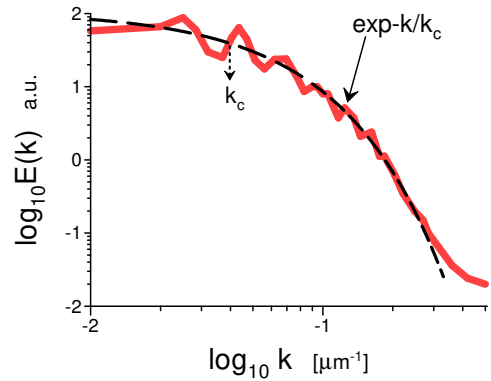} \vspace{-4.1cm}
\caption{Radial power spectrum of a passive scalar measured in a laser-produced laboratory plasma at the chaotic stage.} 
\end{figure}

   It is well known that deterministic chaos for the systems with smooth trajectories is usually characterized by exponential spectrum \cite{fm}-\cite{kds}. Therefore, the energy spectrum observed in Fig. 2 can be considered an indication of deterministic chaos.\\ 
   
   In a recent paper Ref. \cite{rig} results of an experiment with a laser-produced laboratory plasma were reported. At the end of the experiment ($80$ns after the main laser impulse) the spatial structures characteristic to the Rayleigh–Taylor instability are already lost and a highly chaotic state of plasma motion is established. Figure 3 shows a radial power spectrum of a passive scalar (bromine concentration) obtained for this situation. The spectral data were taken from Fig. 4 of the Ref. \cite{rig}. The dashed curve indicates the exponential spectrum Eq. (4) (deterministic chaos).\\
   
   In a recent paper Ref. \cite{mb} results of highly resolved large-eddy simulations (LES) of the transition to turbulence in a Rayleigh-Taylor mixing layer between two incompressible,
miscible fluids were reported. In order to study different stages of the transition the variety of the large eddy simulations were performed for different values of the grid Grashof number: $G_{\Delta} = 2gA \Delta^3/ \nu^2$ (where $\Delta$ is the local mesh spacing of the computational grid). The periodic boundary conditions were used in the horizontal dimensions $x$ and $y$, and the non-penetrating boundary conditions were used on the bottom and top walls ($z = -2/\pi$ and $z = 2\pi$). Simulations were finished when the mixing layer width $h = \pi$. Random initial perturbations to the interface between the two resting fluids were applied. 

  Figure 4 shows the horizontal kinetic energy spectrum averaged over x-y planes between $-\pi/16 \leq z \leq \pi/16$ vs horizontal wavenumbers at the end of the LES. The spectral data were taken from Fig. 5 of the Ref. \cite{mb}. The dashed lines indicate (from bottom to top - increasing $G_{\Delta}$) the exponential (deterministic chaos), the stretched exponential (distributed chaos), and power law $E(k) \propto k^{-5/3}$ (Kolmogorov-like turbulence), cf Sections VIII, IX. 
   
\begin{figure} \vspace{-1.3cm}\centering \hspace{-1cm}
\epsfig{width=.52\textwidth,file=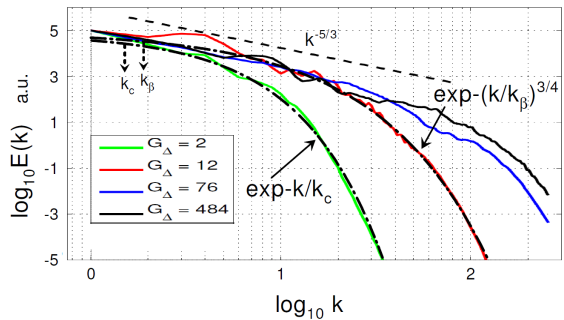} \vspace{-6.3cm}
\caption{Horizontal kinetic energy spectrum averaged over x-y planes between $-\pi/16 \leq z \leq \pi/16$ vs horizontal wavenumbers at the end of the LES.}
\end{figure}
\begin{figure} \vspace{-0.5cm}\centering \hspace{-1cm}
\epsfig{width=.53\textwidth,file=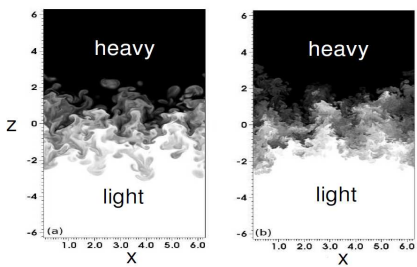} \vspace{-8.4cm}
\caption{Slice (2D contours) of the mass fraction of the heavy fluid $(\rho -\rho_1)/(\rho_2-\rho_1)$ for Grashof numbers (a) $G_{\Delta} = 12$ (distributed chaos) and (b) $G_{\Delta} = 484$ (Kolmogorov-like turbulence), cf Fig. 4.} 
\end{figure}
  
   Figure 5 (adapted from Fig. 6 of the Ref. \cite{mb}) shows a slice (2D contours) of the mass fraction of the heavy fluid $(\rho -\rho_1)/(\rho_2-\rho_1)$ for Grashof numbers (a) $G_{\Delta} = 12$ (distributed chaos) and (b) $G_{\Delta} = 484$ (Kolmogorov-like turbulence), cf Fig. 4. 
   
   From Figs. 4 and 5 one can see that the Kolmogorov-like turbulent stage exhibits significantly finer scales than those observed at the distributed chaos stage. \\

\section{Spontaneous breaking of local reflectional symmetry}   
   
    The global reflection symmetry of a flow (and, consequently, related to this symmetry zero global helicity) does not always mean that also the point-wise helicity is identically equal to zero. Moreover,  there can occur a spontaneous breaking of the local reflectional symmetry due to the spontaneous helicity fluctuations in the chaotic/turbulent flows, and the vorticity blobs (moving with the fluid) with non-zero sign-definite blob's helicity can appear \cite{moff1}-\cite{bkt}. To preserve initial zero global helicity (global reflectional symmetry) these positive and negative helicities localized in the blobs should be canceled at the global (overall) average. \\
    
     Let us consider this system of the vorticity blobs moving with the fluid. Vorticity is tangential: ${\boldsymbol \omega} \cdot {\bf n}=0$, at the boundaries of the vorticity blobs (here ${\boldsymbol \omega}$ is vorticity field). \\
 
  The helicity in the blob with spatial volume $V_j$ is
$$
H_j = \int_{V_j} h({\bf r},t) ~ d{\bf r}.  \eqno{(5)}
$$
where  $j$ denotes the number of the corresponding blob, and $h({\bf r},t) = {\bf v} \cdot {\boldsymbol \omega}$ is the helicity distribution.\\

   The moments of the helicity distribution are \cite{lt} ,\cite{mt}
$$
{\rm I_n} = \lim_{V \rightarrow  \infty} \frac{1}{V} \sum_j H_{j}^n  \eqno{(6)}
$$
where $V$ denotes the total volume of the blobs.\\

 The global helicity ${\rm I_1}$ and odd moments are identically equal to zero due to the global reflectional symmetry. \\
 
   The global helicity ${\rm I_1}$ is not an ideal (nondissipative) invariant of the system (1-3) due to the buoyancy forces. However, the buoyancy forces (generating the motion) affect mainly the {\it large} spatial scales \cite{cz} while for high moments the vorticity blobs with small characteristic scales provide the main contribution to the sum Eq. (6)  \cite{bt}. Therefore, the high moments can be treated as adiabatic invariants in this case. Moreover, the spontaneous appearance of the small-scale vorticity blobs with non-zero (sign-definite) helicity naturally should be much more probable than the spontaneous appearance of the large-scale vorticity blobs with non-zero (sign-definite) helicity.  \\

   If we denote the helicity of the vorticity blobs having negative kinetic helicity as $H_j^{-}$, and those having positive helicity as $H_j^{+}$, then we can denote
$$
{\rm I_n^{\pm}} = \lim_{V \rightarrow  \infty} \frac{1}{V} \sum_j [H_{j}^{\pm}]^n  \eqno{(7)}
$$ 
where the summation in Eq. (7) takes into account the vorticity blobs with negative (or positive) helicity only.  \\

 The odd moments  ${\rm I_n} = {\rm I_n^{+}} + {\rm I_n^{-}} =0$ (due to the global reflectional symmetry), then ${\rm I_n^{+}} = - {\rm I_n^{-}}$, and the non-zero higher odd moments $|{\rm I_n^{\pm}}|$ can be treated as adiabatic-invariants for the case of spontaneously broken local reflectional symmetry. \\
  
\section{Introduction to helical distributed chaos}

An ensemble averaging should be used for the fluctuating characteristic scale $k_c$ in the exponential spectrum Eq. (4)
$$
E(k) \propto \int_0^{\infty} P(k_c) \exp -(k/k_c)dk_c \eqno{(8)}
$$    
where $P(k_c)$ is a probability distribution of the $k_c$. Of course, corresponding chaos is not deterministic, but the trajectories can be still smooth and the corresponding spectrum will be stretched exponential (we will call this situation `distributed chaos'). In order to check this let us consider the case when the third moment of the helical distribution $|{\rm I}_3^{\pm}|$ can be treated as an adiabatic-invariant. 
  
  In this case, the moments ${\rm I}_n^{\pm}$ with $n > 3$ can be also treated as adiabatic-invariants. However, ${\rm I}_ n^{\pm}$ with a minimal value $n_{min}$ high enough to be treated as an adiabatic invariant has the maximal probability to determine the behavior of the chaotic flow, since the attractor's basin of this moment is thicker than the attractor's basins of other moments having $n > n_{min}$.\\

\begin{figure} \vspace{-1.45cm}\centering
\epsfig{width=.47\textwidth,file=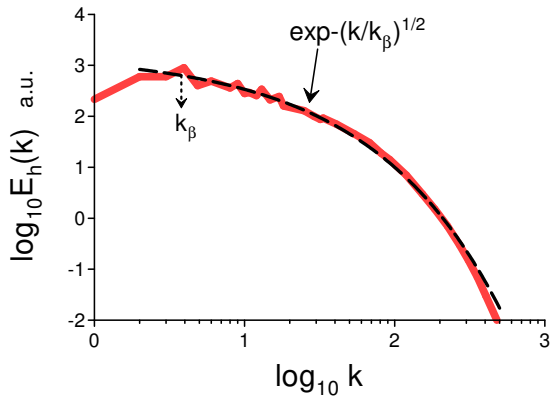} \vspace{-4.4cm}
\caption{ Power spectrum of horizontal velocity vs the horizontal wavenumber $k = \sqrt{k_x^2 + k_y^2}$ computed over the midplane of the mixing layer at $A = 0.5$.} 
\end{figure}
\begin{figure} \vspace{-1.04cm}\centering
\epsfig{width=.46\textwidth,file=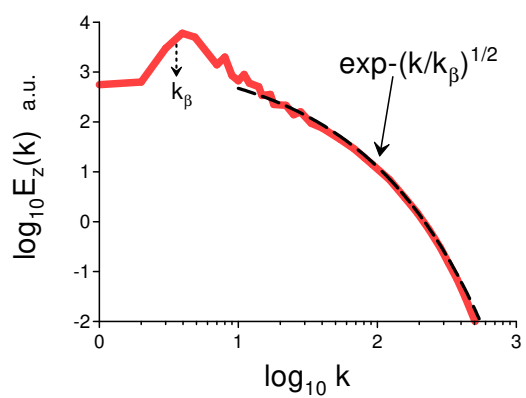} \vspace{-4.55cm}
\caption{ As in Fig. 6 but for vertical velocity. } 
\end{figure}
\begin{figure} \vspace{-0.5cm}\centering
\epsfig{width=.47\textwidth,file=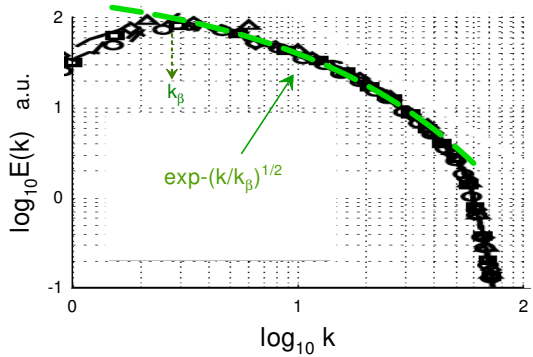} \vspace{-4.49cm}
\caption{ Horizontal kinetic energy spectrum at $A = 0.05$. } 
\end{figure}
     
   The probability distribution of the characteristic velocity $v_c$ and dimensional considerations can be used to find $P(k_c)$ in Eq. (8) for the distributed chaos dominated by $|I_3^{\pm}|$. Indeed, it follows from the dimensional considerations that
$$
 v_c \propto |I_3^{\pm}|^{1/6}~ k_c^{1/2}    \eqno{(9)}
$$       
 
   Using Eq. (9) and assuming a normal distribution of the characteristic velocity $v_c$  \cite{my} one can readily obtain the probability distribution $P(k_c)$  
$$
P(k_c) \propto k_c^{-1/2} \exp-(k_c/4k_{\beta})  \eqno{(10)}
$$
were $k_{\beta}$ is a constant parameter.\\

    Substitution of the $P(k_c)$ from the Eq. (10) into the Eq. (8) results in a stretched exponential spectrum
$$
E(k) \propto \exp-(k/k_{\beta})^{1/2}.  \eqno{(11)}
$$

    The above and further consideration is applicable for both incompressible and compressible fluids \cite{moff1},\cite{mt}.  It should be also noted that the results obtained for the helical distributed chaos can be extended on a dissipative range of scales as well (see the end of the Section VIII).\\
 
\begin{figure} \vspace{-1.3cm}\centering
\epsfig{width=.47\textwidth,file=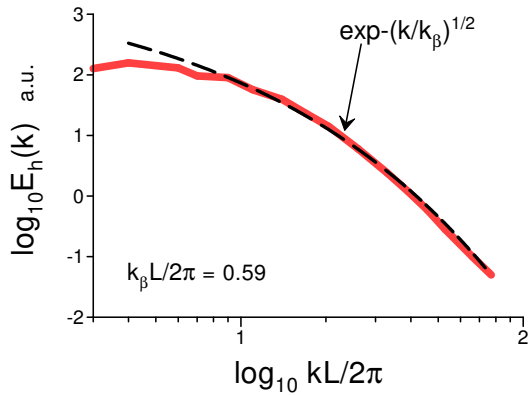} \vspace{-4.35cm}
\caption{Horizontal kinetic energy spectra vs the horizontal wavenumber computed at the midplane of the mixing layer.} 
\end{figure}
\begin{figure} \vspace{-0.5cm}\centering
\epsfig{width=.46\textwidth,file=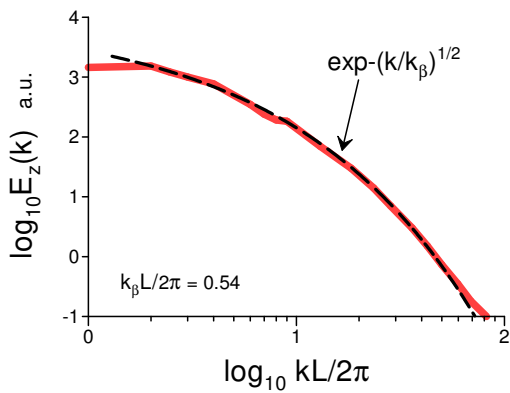} \vspace{-4cm}
\caption{ As in Fig. 9 but for vertical kinetic energy. } 
\end{figure}
\begin{figure} \vspace{-1.3cm}\centering
\epsfig{width=.46\textwidth,file=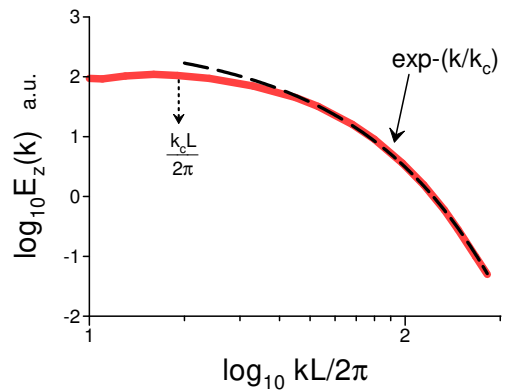} \vspace{-4.1cm}
\caption{ As in Fig. 9 but for vertical magnetic energy. } 
\end{figure}
  
  Figures 6 and 7 show the power spectra of horizontal and vertical velocity vs the horizontal wavenumber $k = \sqrt{k_x^2 + k_y^2}$ computed at the midplane of the mixing layer. The spectral data were taken from Fig. 9 of Ref. \cite{cz}. The DNS for two miscible fluids was performed in a cubic domain with free-slip and impermeable bottom and top boundaries, and periodic boundary conditions in the two horizontal directions. The horizontal separating interface was randomly perturbed to initiate the Rayleigh-Taylor instability. The `incompressibility' condition on the dilatation was taken as
 $$     
 \nabla \cdot {\bf u} = \nabla \cdot \rho D \nabla \rho^{-1}   \eqno{(12)}
 $$
 where $D$ is the Fickian diffusivity. The Schmidt number $Sc = \nu/D$ was taken equal to 1, and the Atwood number $A = 0.5$. The horizontal and vertical Taylor-Reynolds numbers corresponding to the spectra shown in Figs. 6 and 7 were $R_{\lambda} = 65$ and $224$. At this stage of the DNS, the dilatation was found to be negligible. 
 
 The dashed lines in Figs. 6 and 7 indicate the stretched exponential spectrum Eq. (11) (the distributed chaos).  The position of the $k_{\beta}$ in these figures indicates that the distributed chaos was determined by the large-scale coherent structures (related to the spectral peaks) in this case. \\
 
  The previous example corresponds to a comparatively large Atwood number $A =0.5$. Let us now consider an example for a small Atwood number $A = 0.05$. In a recent paper Ref. \cite{morgan} results of a large-eddy simulation for two incompressible, miscible fluids were reported. The simulation was performed in a rectangular cuboid with free-slip and impermeable bottom and top boundaries, and periodic boundary conditions in the two horizontal directions. Initial random perturbations to the interface were used to initiate the Rayleigh-Taylor instability. \\
  
    Figure 8 shows the horizontal kinetic energy spectrum averaged over all horizontal planes where $4 \overline{\rm Y_HY_L} \geqslant 0.7$ (${\rm Y_H} =(\rho- \rho_L)/(\rho_H -\rho_L)$ is the mass fraction of the heavy fluid and ${\rm Y_L} = 1 - {\rm Y_H}$). The different symbols correspond to different mesh resolutions. The spectral data were taken from Fig. 3a of the Ref. \cite{morgan}. The dashed line in Fig. 8 indicates the stretched exponential spectrum Eq. (11) (the distributed chaos). The position of the $k_{\beta}$ in this figure indicates that the distributed chaos was determined by the large-scale coherent structures (related to the spectral peak) in this case as well. \\

   An interesting example was considered in a recent paper Ref. \cite{scout}. In this paper results of a DNS of small-scale magnetohydrodynamic kinematic dynamo generated by the Rayleigh-Taylor-induced mixing were reported. At the small-scale (without mean magnetic field) kinematic dynamo the generated from small seeds magnetic field has negligible effects on the hydrodynamics. The dynamics of the magnetic field ${\bf B}$ was described by the equation
$$
\frac{\partial {\bf B}}{\partial t} = \nabla \times ( {\bf u} \times
    {\bf B}) +\eta \nabla^2 {\bf B}    \eqno{(13)} 
$$
  where $\eta$ is the resistivity.\\
  
  The DNS for two electrically conducting fluids was performed in a cubic domain with reflecting, free-slip and impermeable bottom and top boundaries, and periodic boundary conditions in the two horizontal directions. The vertical velocity was randomly perturbed to initiate the Rayleigh-Taylor instability. The Prandtl number $Pr =1$ and the magnetic Prandtl number $Pm = \nu /\eta =3$, $A= 0.67$, $g = 0.65$, and $\nu^{-1} = 1\times 10^5$. \\
  
  Figures 9 and 10 show the horizontal and vertical kinetic energy spectra vs the horizontal wavenumber computed at the midplane of the mixing layer ($L$ is the size of the cubic domain). The spectral data were taken from Fig. 7 of the Ref. \cite{scout}. The dashed lines in Figs. 9 and 10 indicate the stretched exponential spectrum Eq. (11) (the distributed chaos). Figure 11 shows the corresponding vertical magnetic energy spectrum. The spectral data were taken from Fig. 7 of the Ref. \cite{scout}. The dashed line in Fig. 11 indicates the exponential spectrum Eq. (4) (the deterministic chaos). 
   
     One can see that in this case, the distributed chaos in the Rayleigh-Taylor-induced mixing generates a deterministically chaotic magnetic field (cf. Refs. \cite{b1},\cite{b2}).
     
\begin{figure} \vspace{-1.3cm}\centering
\epsfig{width=.45\textwidth,file=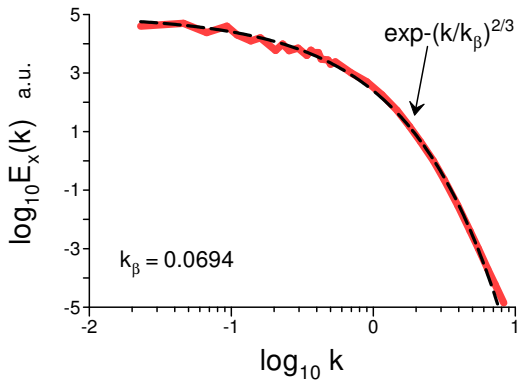} \vspace{-4.05cm}
\caption{ Midplane power spectrum of the $x$ velocity components vs the horizontal wavenumber.} 
\end{figure}
\begin{figure} \vspace{-0.45cm}\centering
\epsfig{width=.45\textwidth,file=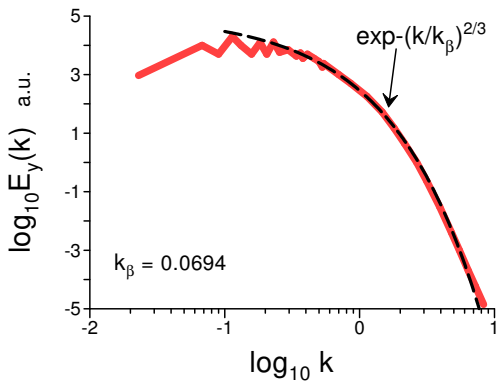} \vspace{-4.3cm}
\caption{ As in Fig. 12 but for $y$ velocity component. } 
\end{figure}
\begin{figure} \vspace{-1.75cm}\centering
\epsfig{width=.46\textwidth,file=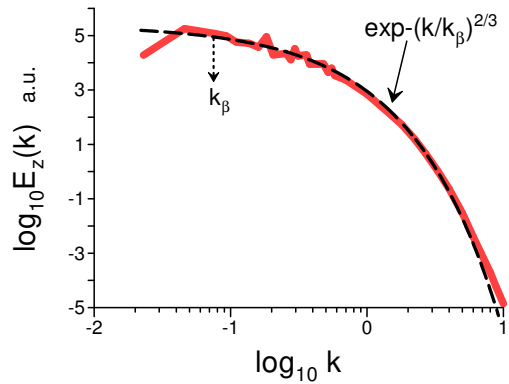} \vspace{-3.85cm}
\caption{ As in Fig. 12 but for $z$ (vertical) velocity component. } 
\end{figure}
   
\section{The helical distributed chaos in more detail}    
  
 In the general case, the estimation Eq. (9) can be replaced by
$$
 v_c \propto  |I_n^{\pm}|^{1/2n}~ k_c^{\alpha_n}   \eqno{(14)}
 $$  
 for odd moments and by 
 $$
  v_c \propto  I_n^{1/2n}~ k_c^{\alpha_n},    \eqno{(15)}
 $$  
 for even moments,  where
$$
\alpha_n = 1-\frac{3}{2n},  \eqno{(16)}
$$  
and the stretched exponential spectrum Eq. (11) in the case of smooth dynamics can be generalized as
$$
E(k) \propto \int_0^{\infty} P(k_c) \exp -(k/k_c)dk_c \propto \exp-(k/k_{\beta})^{\beta} \eqno{(17)}
$$  
 
  Then for large $k_c$ the distribution $P(k_c)$ can be estimated from Eq. (17) \cite{jon}
$$
P(k_c) \propto k_c^{-1 + \beta/[2(1-\beta)]}~\exp(-\gamma k_c^{\beta/(1-\beta)}) \eqno{(18)}
$$     
(here the $\gamma$ is a constant).\\

    If $v_c$ has normal distribution \cite{my} one can readily obtain a relationship between the exponents $\beta_n$ and $\alpha_n$  from the Eqs. (14-15) and (18)
$$
\beta_n = \frac{2\alpha_n}{1+2\alpha_n}  \eqno{(19)}
$$
 
  Substituting Eq. (16) into the Eq. (19) one obtains
 $$
 \beta_n = \frac{2n-3}{3n-3}   \eqno{(20)}  
 $$

  Then for $n \gg 1$
$$
E(k) \propto \exp-(k/k_{\beta})^{2/3}  \eqno{(21)}
$$ 
and for $n=2$ (the Levich-Tsinober invariant \cite{lt})
$$
E(k) \propto \exp-(k/k_{\beta})^{1/3}  \eqno{(22)}
$$ 
   
   Let us recall that the above consideration is valid for both incompressible and compressible cases.

\begin{figure} \vspace{-1.3cm}\centering \hspace{-1.2cm}
\epsfig{width=.53\textwidth,file=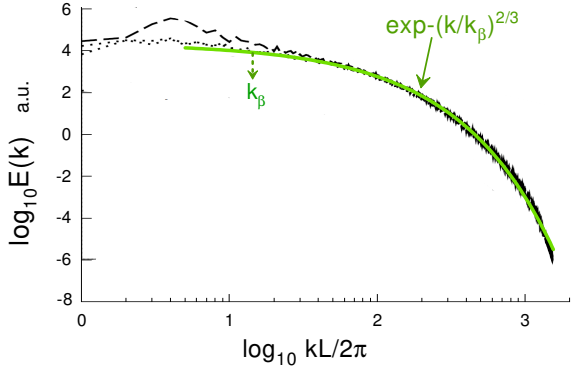} \vspace{-5.63cm}
\caption{ The midplane power spectra of the horizontal (dotes) and vertical (dashes) velocities. } 
\end{figure}

\begin{figure} \vspace{-0.5cm}\centering
\epsfig{width=.445\textwidth,file=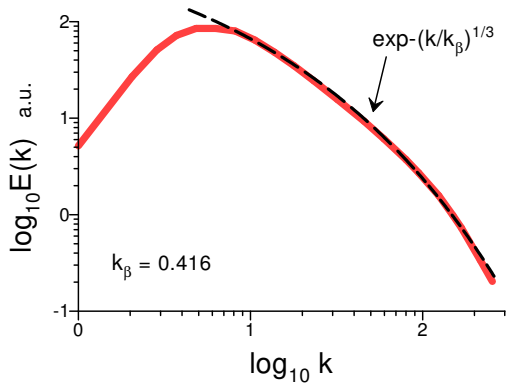} \vspace{-3.92cm}
\caption{Power spectrum of 3D velocity field.} 
\end{figure}
\begin{figure} \vspace{-1cm}\centering
\epsfig{width=.445\textwidth,file=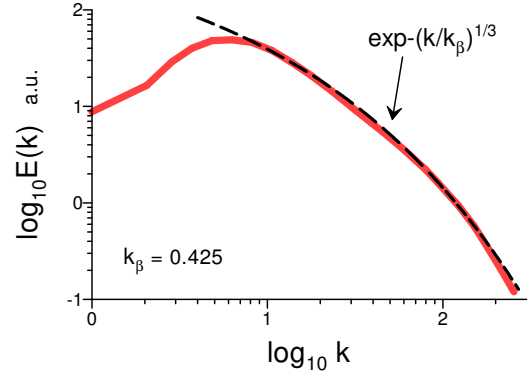} \vspace{-3.95cm}
\caption{Kinetic energy spectrum.} 
\end{figure}
    
 \section{Numerical simulations and laboratory experiments}
 
   In paper Ref. \cite{mu} results of a direct numerical simulation of Rayleigh-Taylor-induced mixing of the incompressible, variable-density fluids in a rectangular cuboid ($L_x \times L_y \times L_z = 28.8 \times 18 \times 24$)  at a small Atwood number $A = 7.5 \times 10^{-4}$, and $Pr = Sc = 7$, were reported. At the end of the simulation $Re_h \simeq 1700$ (based on the vertical size $h$ of the mixing region). The free-slip and impermeable bottom and top boundaries were considered and periodic boundary conditions were imposed in the two horizontal directions. Initial random perturbations to the interface  and velocity field were used to initiate the Rayleigh-Taylor instability.\\
   
    Figures 12, 13, and 14 show the midplane power spectra of the velocity components vs the horizontal wavenumber. The spectral data were taken from Fig. 3.20 of the Ref. \cite{mu}. The dashed lines in these figures indicate the stretched exponential spectrum Eq. (21) (the distributed chaos).\\
    
    Results of an analogous simulation were reported in Ref. \cite{cc}. The simulation was performed in a cubic (at the end of the simulation) box at $A = 0.5$, $Sc =1$, and $Re=32000$. Figure 15 shows the power spectra of the horizontal (dotes) and vertical (dashes) velocities. The spectral data were taken from Fig. 2 of the Ref. \cite{cc}.  The solid curve indicates the stretched exponential spectrum Eq. (21) (the distributed chaos).\\

    In a recent paper Ref. \cite{zhao} results of a direct numerical simulation of a single-species two-density Rayleigh-Taylor model governed by the fully compressible Navier-Stokes equations and the ideal gas equation of state with the following parameters: $A = 0.5$, $Gr = 15.07$, $Re = 13854$, $Ma =0.45$, $Pr =1$, and $g = 1$ were reported. The DNS was performed in a rectangular cuboid $L_x=L_y = 3.2$, $L_z = 2L_x$ with the standard boundary conditions (see above). 
    
    Figure 16 shows the power spectrum of 3D velocity field and figure 17 shows the kinetic energy spectrum (taking into account fluctuations of the density field). The spectral data were taken from Figs. 5e and 5f of the Ref. \cite{zhao}. One can see that the spectra are rather similar. The dashed lines in these figures indicate the stretched exponential spectrum Eq. (22) (the distributed chaos). \\

    In paper Ref. \cite{zin} role of the Rayleigh-Taylor instability-induced mixing in accelerating a thermonuclear flame in Type Ia supernovae had been studied using a single density low–Mach number inviscid Navier-Stokes numerical simulation with an advection projection reaction formalism (with the fuel half-oxygen and half-carbon and only the carbon is burned). The spatial domain $L_x =L_y =53.5$ and $L_z = 107.0$ has been used and $A = 0.28$, $g =10^9$. The boundary conditions in $x$ and $y$ directions are periodic. At the bottom boundary fuel was entered at the laminar flame speed and the top boundary is outflow. The instability is triggered by random perturbations. Rayleigh-Taylor unstable flame enters a distributed burning regime near the end of the numerical simulation.\\ 
    
    Figures 18-21 show the transition from the smooth mixing regime (the distributed chaos at $t =8.11\times 10^{-4}$) to the nonsmooth regime (Kolmogorov-like turbulence at $t = 1.1 \times 10^{-3}$) at the late times of the simulation. Figure 18 was adapted from Figs. 2c and 2f of the Ref. \cite{zin} and the spectral data for the kinetic energy spectra (after projecting out the compressible part) were taken from Figs. 2c, 2d, and 2f of the Ref. \cite{zin}. The dashed lines indicate the stretched exponential spectra Eqs. (21-22) (the distributed chaos). The straight solid line indicates the Kolmogorov-like spectrum $E(k) \propto k^{-5/3}$. In the Fig. 21 one can see the appearance of the Kolmogorov-like inertial range at the large scales (small $k$) and the appearance of the sub-inertial range dominated by the distributed chaos Eq. (21) at the small scales (large $k$). \\

\begin{figure} \vspace{-1.4cm}\centering \hspace{-2.5cm}
\epsfig{width=.61\textwidth,file=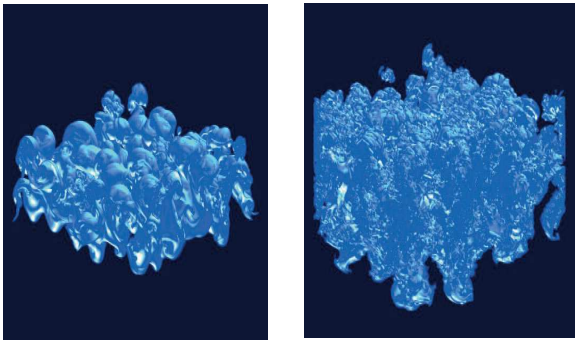} \vspace{-7.6cm}
\caption{Volume rendering of the mass fraction (carbon). Left at $t =8.11 \times 10^{-4}$ and right at  $t = 1.1 \times 10^{-3}$.}
\end{figure}
\begin{figure} \vspace{-0.5cm}\centering
\epsfig{width=.45\textwidth,file=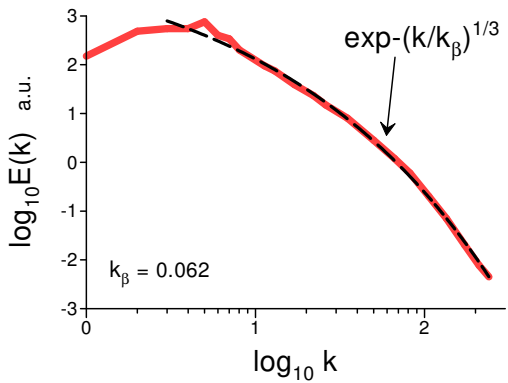} \vspace{-4.35cm}
\caption{Kinetic energy spectrum (after projecting out the compressible part) at $t =8.11 \times 10^{-4}$.} 
\end{figure}
\begin{figure} \vspace{-1cm}\centering
\epsfig{width=.45\textwidth,file=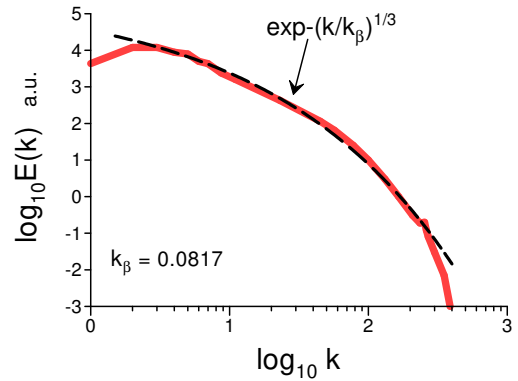} \vspace{-4.23cm}
\caption{As in Fig. 19 but at $t =9.43 \times 10^{-4}$. } 
\end{figure}
\begin{figure} \vspace{-0.6cm}\centering
\epsfig{width=.45\textwidth,file=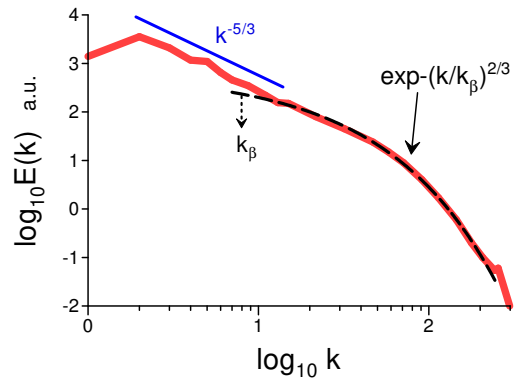} \vspace{-3.91cm}
\caption{As in Fig. 19 but at $t = 1.1 \times 10^{-3}$.} 
\end{figure}

    In paper Ref. \cite{kraft} results of laboratory experiment measurements of Rayleigh-Taylor-induced mixing in an air/helium channel were reported. Heavy (air) and light (a helium/air mixture) fluids entered a channel and were separated by a splitter plate. At the end of the plate, the two gas streams could mix and form a Rayleigh-Taylor mixing layer. The vertical velocity fluctuations dominated this process at the late time. The measurements of the velocity fluctuations were made by a hot-wire probe at the centerline (midplane) of the mixing layer. \\

    Figure 22 shows the power spectrum of the measured vertical velocity fluctuations at $Re_h = 1450$ and $A = 0.03$. The spectral data were taken from Fig. 4.11 of the Ref. \cite{kraft}. $H$ is the height of the channel. The dashed line indicates the stretched exponential spectrum Eq. (22) (the distributed chaos).  \\ 
 
   In paper Ref. \cite{aku} results of a rather similar experiment were reported (case A1S0). Figure 23 shows power spectra of streamwise (lower curve) and cross-stream (upper curve) velocity fluctuations at the late time of the development of the Rayleigh-Taylor-induced mixing layer for $A = 0.035$ and $Re_D = 34000$ ($Re_D$ is the Reynolds number constructed using the hydraulic diameter $D$ of the test section). The spectral data were taken from Fig. 22e of the Ref. \cite{aku}. The black dashed line indicates the stretched exponential spectrum Eq. (22) (the distributed chaos, cf Ref. \cite{pa},\cite{as},\cite{b3}).\\

  Finally, let us briefly discuss the results of implicit large-eddy simulations of the Richtmyer-Meshkov instability-induced mixing after reshock reported in paper Ref. \cite{thorn}. The Richtmyer-Meshkov instability is a result of an impulsive acceleration (usually caused by an incident shock wave) of the interface separating two fluids with different densities (the Richtmyer-Meshkov instability can be referred as shock-induced Rayleigh-Taylor instability). The Kelvin-Helmholtz instability of the shear layers along the interface strongly enhances the mixing in this case. Often a reshock causes a second impulsive acceleration, that results in an additional intensification of the mixing. \\
  
   The simulation is based on the two-component Euler equations (without viscosity) and the ideal gas equation of state for both fluids with non-reflective boundary conditions. In the two homogeneous directions the periodic boundary conditions have been used. In the perpendicular direction an elongated domain was chosen. The broadband initial perturbation of the separating surface was typical for inertial confinement fusion capsules. The post-shock $A = 0.49$. \\
   
   Figure 24 shows kinetic energy spectra, computed in the centre of the mixing layer using 2D-slices in the homogeneous directions, vs the `homogeneous' wave number. The spectral data were taken from Fig. 15 of the Ref. \cite{thorn} for the broadband case. The dashed curves indicate the stretched exponential spectra Eqs. (11) and (22) (the distributed chaos).\\

\section{Unstably stratified homogeneous turbulence}    
 
\begin{figure} \vspace{-1.4cm}\centering
\epsfig{width=.45\textwidth,file=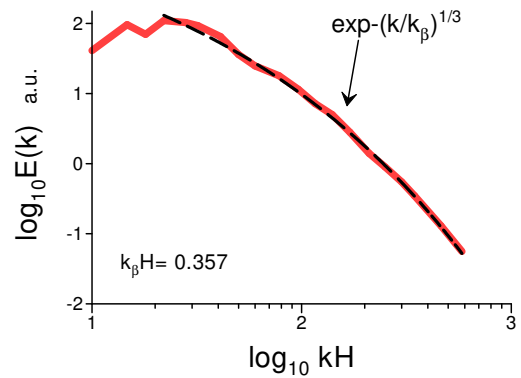} \vspace{-4.05cm}
\caption{Power spectrum of the measured vertical velocity fluctuations at $Re_h = 1450$ and $A = 0.03$.} 
\end{figure}
\begin{figure} \vspace{-0.5cm}\centering
\epsfig{width=.45\textwidth,file=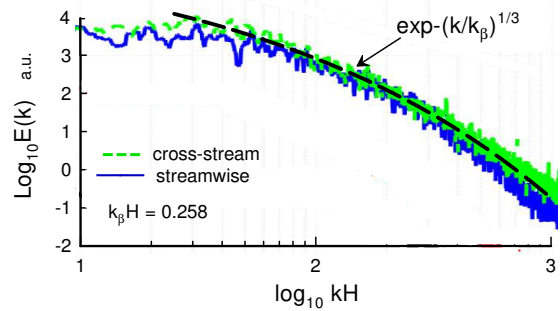} \vspace{-4.95cm}
\caption{Power spectra of streamwise (lower curve) and cross-stream (upper curve) velocity fluctuations at $Re_D = 34000$ and $A = 0.035$. } 
\end{figure}
\begin{figure} \vspace{-0.45cm}\centering
\epsfig{width=.5\textwidth,file=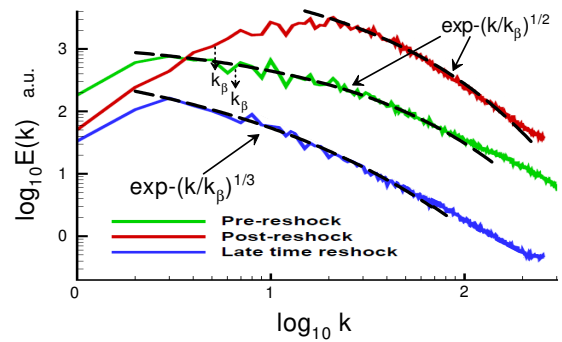} \vspace{-6cm}
\caption{Kinetic energy spectra for the broadband case. } 
\end{figure}

  Under certain conditions, the Rayleigh-Taylor-induced chaotic/turbulent mixing can be modeled by the mixing in unstably stratified homogeneous flows (see, for instance, Ref. \cite{bur} and references therein). In the paper Ref. \cite{bur} such mixing was studied using the Navier–Stokes–Boussinesq equations with a destabilizing background density gradient:
 $$ 
\frac{ \partial{\mathbf u}}{\partial t} +{\mathbf u} \cdot \nabla {\mathbf u}  = -\frac{1}{\rho_{0}} \nabla P + N \theta  {\bf e_g} + \nu \nabla^2 {\mathbf u}  \eqno{(23)}, 
$$ 
$$     
\frac {\partial \theta}{\partial t} +{\mathbf u} \cdot \nabla \theta  = N~ {\bf u} \cdot {\bf e_z} + \kappa \nabla^2 \theta \eqno{(24)},
$$  
$$
 \nabla \cdot {\bf u} = 0 \eqno{(25)};
$$ 
where the fluctuating buoyancy $\theta$ is rescaled as velocity, $\rho_0 = (\rho_{heavy} + \rho_{light})/2$, $N=\sqrt{2Ag\Gamma}$ ($\Gamma$ is the mean density gradient of heavy fluid). 

  The equations were numerically (DNS) solved in a cubic box with triply periodic boundary conditions. The initial conditions were prepared by creating a chaotic (low Reynolds number $Re \simeq 3$, cf Ref. \cite{kds}) homogeneous isotropic motion. Figure 25 shows the kinetic energy spectrum of the motion at the moment ($t = 0$) when the gravity and the mean density gradient of heavy fluid were applied. The spectral data were taken from Fig. 6 of the Ref. \cite{bur}. The dashed curve indicates the exponential spectrum Eq. (4) (deterministic chaos). The application of the buoyancy forces results in the rapid increase of the Reynolds number. \\

    Figures 26 and 27 show kinetic energy spectra obtained in the DNS reported in the Ref. \cite{bur} for $N=4$ and $\nu=\kappa=2.5\times 10^{-3}$ at $t=3$ and $t =6$ (in the terms of the Ref. \cite{bur}). The spectral data were taken from Fig. 6 of the Ref. \cite{bur}. The dashed curves indicate the stretched exponential spectrum Eq. (21) (the helical distributed chaos).\\
    
      In paper Ref. \cite{sg} results of measurements in an oceanic buoyancy-driven mixing layer were reported. Figure 28 shows a typical power spectrum of cross-stream velocity. The spectral data were taken from Fig. A1 of the Ref. \cite{sg}.  The dashed curve indicates the stretched exponential spectrum Eq. (21) (the helical distributed chaos).

\section{Chkhetiani invariants}
  
\begin{figure} \vspace{-1.2cm}\centering
\epsfig{width=.45\textwidth,file=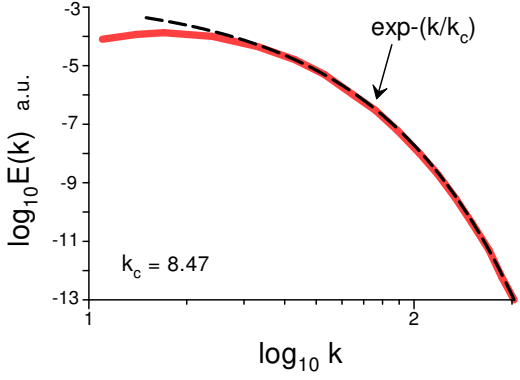} \vspace{-3.85cm}
\caption{Kinetic energy spectrum for $t=0$.} 
\end{figure}
\begin{figure} \vspace{-0.86cm}\centering
\epsfig{width=.45\textwidth,file=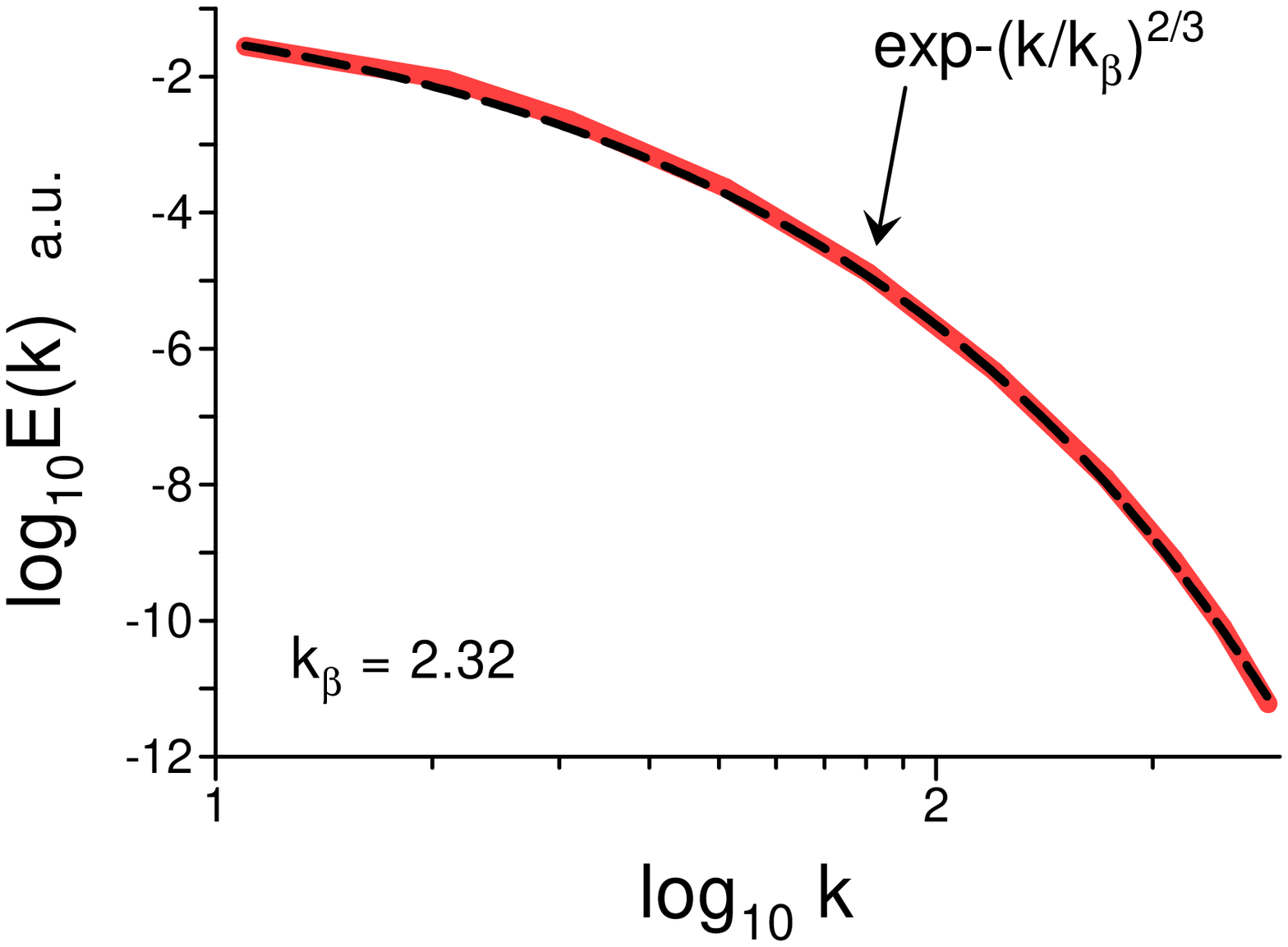} \vspace{-4.2cm}
\caption{Kinetic energy spectrum for $t=3$.} 
\end{figure}
\begin{figure} \vspace{-0.5cm}\centering
\epsfig{width=.45\textwidth,file=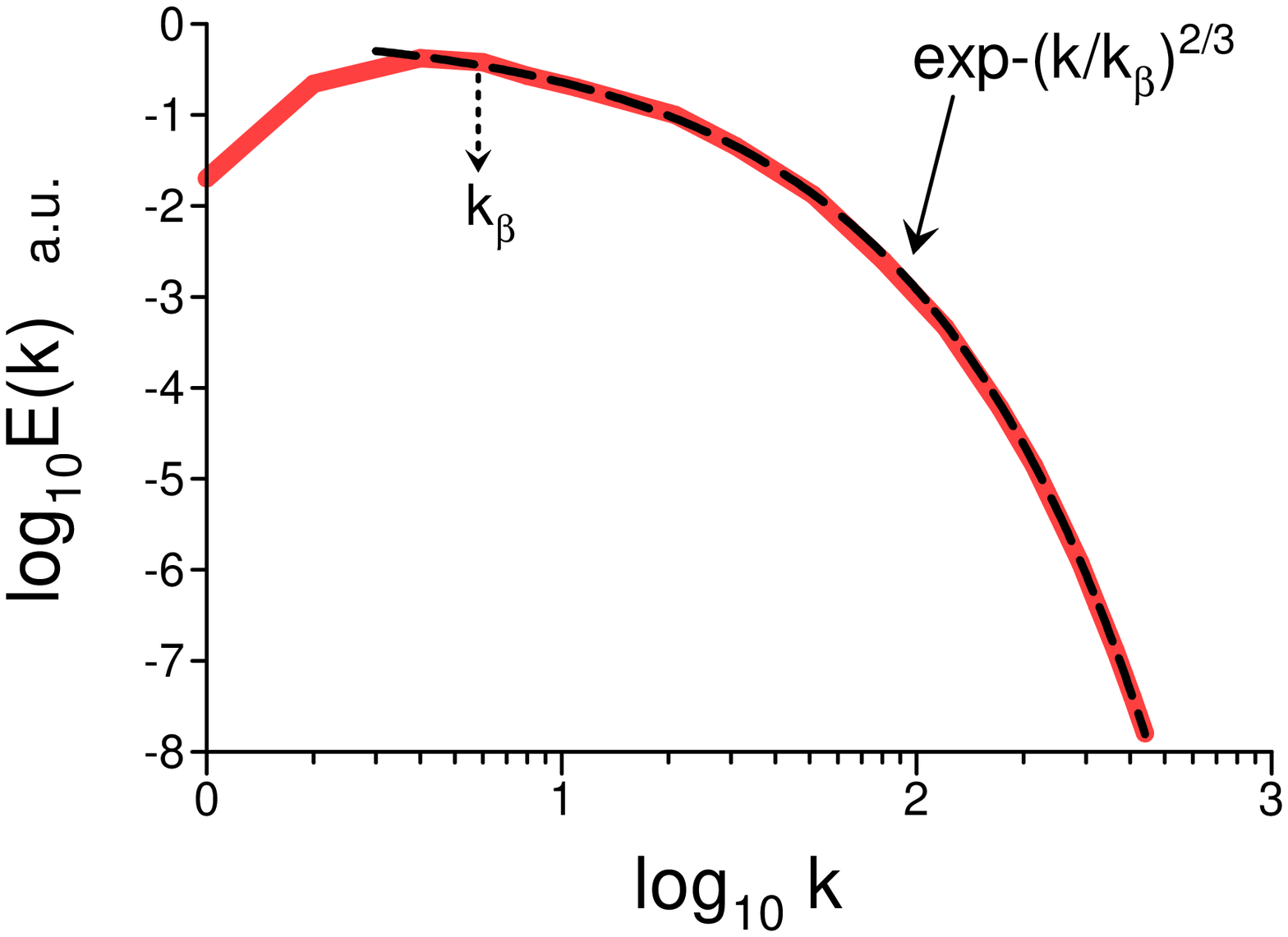} \vspace{-4.13cm}
\caption{Kinetic energy spectrum for $t=6$. } 
\end{figure}
\begin{figure} \vspace{-1.4cm}\centering
\epsfig{width=.48\textwidth,file=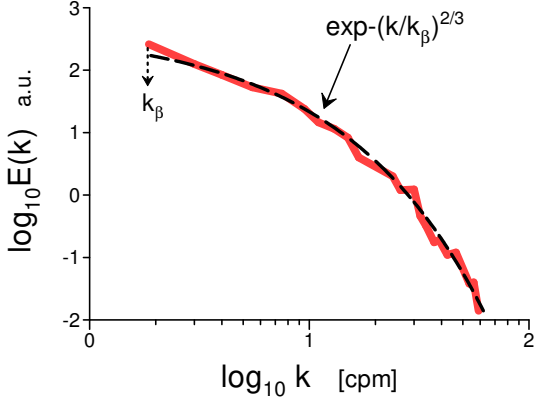} \vspace{-4.6cm}
\caption{A typical power spectrum of cross-stream velocity in an oceanic buoyancy-driven mixing layer.} 
\end{figure}
 
   Energy and helicity conservation takes place in ideal (non-dissipative) fluid dynamics. The Navier-Stokes (dissipative) equations have other fundamental invariants based on the conservation of momentum and angular momentum conservation: Birkhoff-Saffman and Loitsyanskii invariants \cite{my},\cite{bir},\cite{saf},\cite{dav}.  In paper Ref. \cite{otto1} (see also \cite{otto2}) a new group of the fundamental invariants of the Navier-Stokes (dissipative) equations has been added:
 $$
\int r^p\langle {\bf u} ({\bf x},t) \cdot  {\boldsymbol \omega} ({\bf x} + {\bf r}, t) \rangle d{\bf r}   \eqno{(26)}
$$  
 $p =0,1,2$. These invariants were introduced for isotropic homogeneous turbulence. \\
 
  Let us consider the case of spontaneous breaking of the local reflectional symmetry and define 
$$
^p\mathcal{H}_j = \int_{V_j}  r^p\langle {\bf u} ({\bf x},t) \cdot  {\boldsymbol \omega} ({\bf x} + {\bf r}, t) \rangle d{\bf r}  \eqno{(27)}
$$
where $j$ denotes the number of the corresponding vorticity blob. Let us define $^p\mathcal{H}_{j}^{+}$ and $^p\mathcal{H}_{j}^{-}$ depending on its sign.
   
   Then 
 $$
^p\mathcal{I}^{\pm} =   \sum_j~  ^p\mathcal{H}_j^{\pm}    \eqno{(28)}
$$
 where the summation in Eq. (28) takes into account the vorticity blobs with negative (or positive) values of $^p\mathcal H_j$ only. \\
 
 It was already mentioned that the spontaneous appearance of the small-scale vorticity blobs with non-zero (sign-definite) helicity naturally should be much more probable than the spontaneous appearance of the large-scale vorticity blobs with non-zero (sign-definite) helicity. The same is also valid for the vorticity blobs with nonzero $^p\mathcal{H}_j$. Therefore, at certain conditions the small-scale blobs will provide the main contribution to the sum Eq. (28). If for such blobs the $^p\mathcal{H}_j^{\pm}$ can be considered as adiabatic quasi-invariants the $^p\mathcal{I}^{\pm}$ can be also considered as an adiabatic quasi-invariant. \\
 
  In a flow dominated, in a certain range of scales, by the $^0\mathcal{I}^{\pm} $-invariant the dimensional considerations allow the estimation
$$
 v_c \propto  |^0\mathcal{I}^{\pm}|^{1/2}~ k_c,   \eqno{(29)}
$$  
 and following the logic of Section V Eq. (19)
 $$
 \beta = \frac{2\alpha}{1+2\alpha}   \eqno{(30)}
 $$
 we obtain $\beta = 2/3$ (here $\alpha =1$, see Eq. (29)); i.e
 $$
 E(k) \propto \exp(-k/k_{\beta})^{2/3}    \eqno{(31)}
 $$
 In a flow dominated, in a certain range of scales, by the $^1\mathcal{I}^{\pm} $-invariant the dimensional considerations allow the estimation
 $$
 v_c \propto  |^1\mathcal{I}^{\pm}|^{1/2}~ k_c^{3/2},   \eqno{(32)}
$$  
i.e. $\alpha = 3/2$, and from Eq. (30) we obtain
$$
 E(k) \propto \exp(-k/k_{\beta})^{3/4}.    \eqno{(33)}
 $$

   It should be noted that consideration of the moments
$$
{^0\tilde{\mathcal{I}}_n^{\pm}} = \lim_{V \rightarrow  \infty} \frac{1}{V} \sum_j [^0\mathcal{H}_{j}^{\pm}]^n,  \eqno{(34)}
$$
instead of the helicity distribution moments defined by Eq. (7),  allows the extension of the results of Sections III-V on a dissipative range of scales (due to the dissipative character of the Chkhetiani invariants). The summation in Eq. (34) takes into account the vorticity blobs with negative (or positive) values of $^0\mathcal H_j$ only.

\section{Numerical simulations}

\begin{figure} \vspace{-0.95cm}\centering
\epsfig{width=.45\textwidth,file=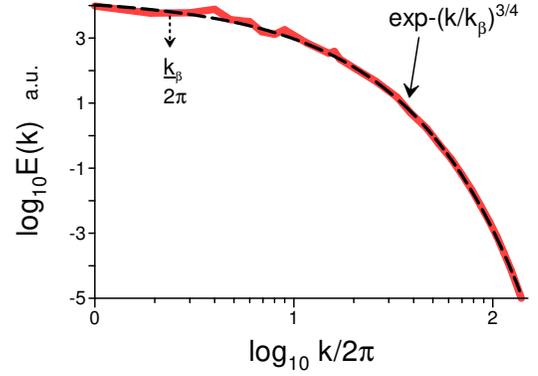} \vspace{-4.cm}
\caption{Kinetic energy spectrum vs horizontal wave number on the mixing layer midplane.  } 
\end{figure}
\begin{figure} \vspace{-0.5cm}\centering
\epsfig{width=.45\textwidth,file=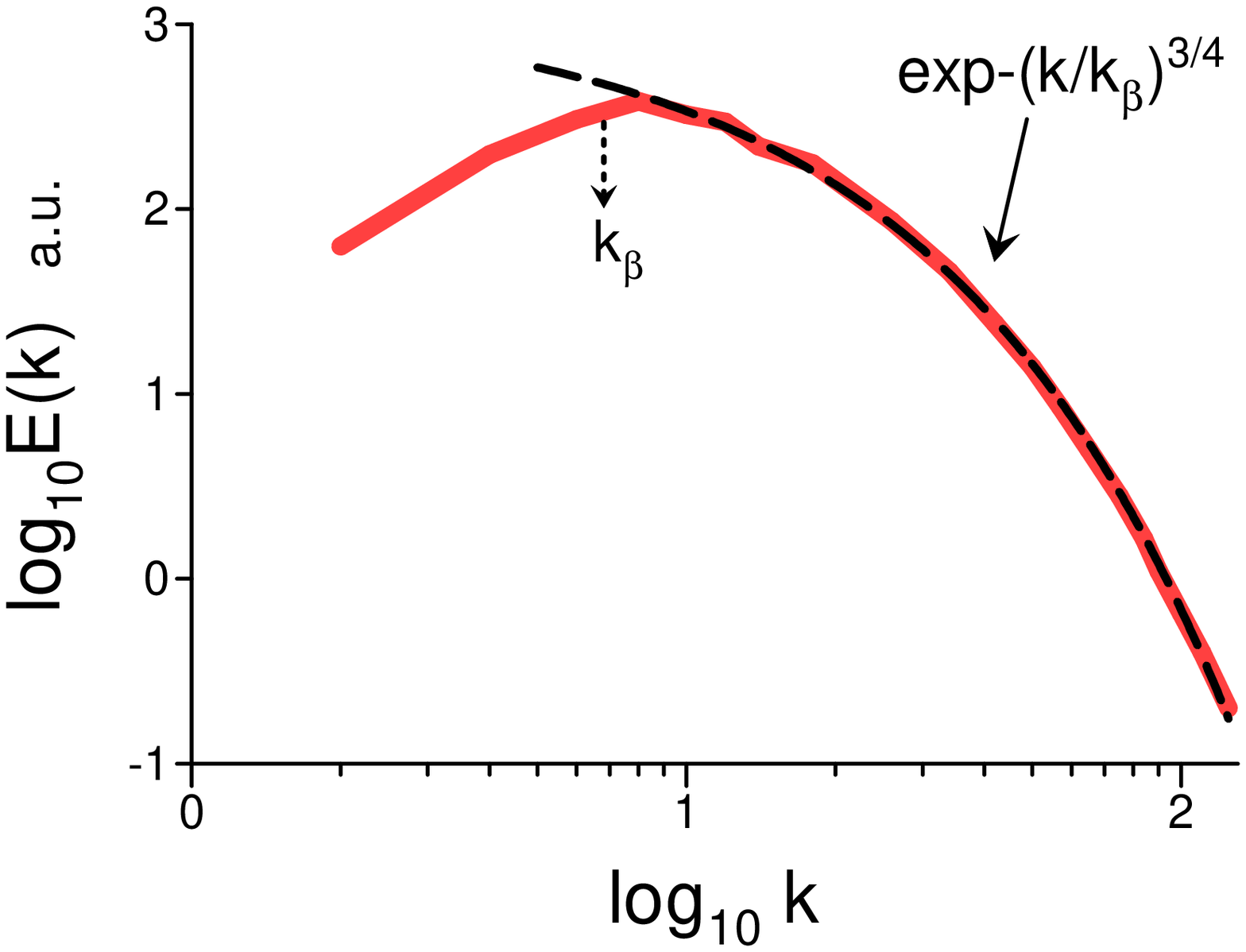} \vspace{-3.7cm}
\caption{ Kinetic energy spectrum at the time $t = \tau$. } 
\end{figure}
\begin{figure} \vspace{-1.2cm}\centering
\epsfig{width=.45\textwidth,file=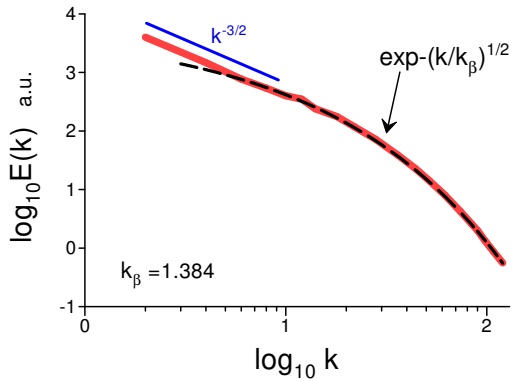} \vspace{-3.87cm}
\caption{Kinetic energy spectrum at the time $t =3.8 \tau$.  } 
\end{figure}

  First of all, let us note that some (and maybe all) above presented numerically calculated spectra with $\beta =2/3$ are representing the spectrum Eq. (31) from Section VIII. Therefore, we will start from the spectra Eq. (33). First example of such a spectrum is shown in Fig. 4 (see description of the corresponding numerical simulation in Section II).  \\
  
  Another example of a numerically computed (DNS) spectrum in the form of Eq. (33) is shown in Fig. 29. The spectral data were taken from Fig. 11 of Ref. \cite{cz2} (the final spectrum). The figure shows the kinetic energy spectrum vs horizontal wave number on the mixing layer midplane. This direct numerical simulation of a Rayleigh-Taylor mixing of two incompressible fluids with no surface tension was performed in a rectangular cuboid with periodic boundary conditions in the horizontal (homogeneous) directions and no-slip conditions on the bottom and top walls at  $A =0.5$ and final $Re_h = 5500$ (based on the vertical size $h$ of the mixing region). The Rayleigh-Taylor instability was triggered by random perturbations of the separating interface, density, and velocity. The dashed curve indicates the stretched exponential spectrum Eq. (33) (the distributed chaos dominated by the $\mathcal{I}_1 ^{\pm}$-invariant).\\

  Results of a small Atwood number direct numerical simulation similar to the first DNS described in Section II (see Fig. 2) were reported in Ref. \cite{bmmv}. In this DNS periodic boundary conditions were used in all three directions ($x,~y,~z$), $Pr =1$, $\alpha g =2$,  $Re_{\lambda} = 196$. \\
  
  Figures 30 and 31 show the kinetic energy spectra computed at the time $t = \tau$ and $t = 3.8 \tau$ (where $\tau = (L_z/Ag)^{1/2}$). The spectral data were taken from Fig. 10a of the Ref. \cite{bmmv}. The dashed curve in Fig. 30 indicates the stretched exponential spectrum Eq. (33) and the stretched exponential spectrum Eq. (11) in Fig. 31. Comparing Figs. 2, 30, and 31 one can see the evolution of the mixing from deterministic chaos to the scaling turbulence via the distributed chaos with the spontaneous breaking of the local reflectional symmetry (cf also Fig. 4).\\

\end{document}